\documentclass[11pt,a4paper]{article}
\usepackage{mathrsfs}
\usepackage{latexsym}
\usepackage{amsfonts}
\usepackage{amssymb}

\newcommand{\be}{\begin{equation}}
\newcommand{\ee}{\end{equation}}
\newcommand{\Gammabol}{\Gamma{}}
\newcommand{\Rbol}{R{}}
\newcommand{\nabol}{\nabla{}}
\newcommand{\onehalf}{{\textstyle{\frac{1}{2}}}}

\def\mt{{\mbox{\tiny{(1)}}}}
\def\my{{\mbox{\tiny{(2)}}}}

\begin{document}
\noindent
{\Large \bf The Nonlinear Essence of Gravitational Waves}
\vskip 0.7cm
\noindent
{\bf R. Aldrovandi, J. G. Pereira and K. H. Vu} \\
{\it Instituto de F\'{\i}sica Te\'orica},
{\it Universidade Estadual Paulista} \\
{\it Rua Pamplona 145},
{\it 01405-900 S\~ao Paulo, Brazil}

\vskip 0.8cm
\noindent
{\bf Abstract~} A critical review of gravitational wave theory is made. It is pointed out that the usual linear approach to the gravitational wave theory is neither conceptually consistent nor mathematically justified. Relying upon that analysis it is then argued that --- analogously to a Yang-Mills propagating field, which must be nonlinear to carry its gauge charge --- a gravitational wave must necessarily be nonlinear to transport its own charge --- that is, energy-momentum. 

%%%%%%%%%%%%%%%%%%%%%%
\section{Introduction}
%%%%%%%%%%%%%%%%%%%%%%
\label{basic}
\setcounter{footnote}{0}

Although they have not yet been directly detected, there are compelling experimental evidences for the existence of gravitational waves, coming basically from the orbital period change of binary pulsar systems~\cite{bina}. These evidences, however, do not give any clue on their correct form. This subject is actually plagued by theoretical difficulties that obscure their physical properties~\cite{cs1}. One of the main problems refers to the concept of energy localization~\cite{energy}. On the one hand, the strong equivalence principle seems to forbid the energy of the gravitational field to be localized~\cite{mtw}. On the other, from the point of view of field theory, it should be possible to define an energy density for the gravitational field. Relying on this view, Synge says in the preface of his classic textbook~\cite{synge} that {\it in Einstein's theory, either there is a gravitational field or there is none, according to as the Riemann tensor does not or does vanish. This is an absolute property; it has nothing to do with any observer's world line}. According to Synge, therefore, the energy of the gravitational field should be localizable independently of any observer. In the same line, Bondi argues that {\it in relativity a non-localizable form of energy is inadmissible, because any form of energy contributes to gravitation and so its location can in principle be found}~\cite{bondi}. This controversial point is the origin of a long-standing discussion on the energy and momentum transported by a gravitational wave.

The study of gravitational waves involves essentially the weak field approximation of Einstein equation, the field equation of general relativity. The geometrical setting of this theory is the tangent bundle:\footnote{We use the Greek alphabet ($\mu, \nu, \rho, \dots = 0, 1, 2, 3$) to denote spacetime indices, and the first half of the Latin alphabet ($a, b, c, \dots = 0, 1, 2, 3$) to denote algebraic indices related to the tangent Minkowski spaces, whose metric is assumed to be $\eta_{ab} = \mbox{diag}~(+1, -1, -1, -1)$.} at each point $x^\mu$ of spacetime, there is attached a Minkowski tangent space with coordinates $x^a = x^a(x^\mu)$. This structure allows a clear distinction between coordinate system and reference frame~\cite{livro}. Given a Lorentz frame
\be
e^a = e^a{}_\mu dx^\mu,
\ee
it is possible to describe it in infinitely many coordinate systems $\{x^\mu\}$. Observe that, although $e^a$ is coordinate independent, under a general spacetime coordinate transformation
\be
x^\mu \to x'^\mu = x^\mu + \xi^\mu(x),
\label{difeo}
\ee 
its components $e^a{}_\mu$ change covariantly:
\be
e^a{}_{\mu'} = \frac{\partial x^\rho}{\partial x^{\mu'}} \, e^a{}_\rho.
\ee
On the other hand, there are infinitely many frames $e^a$, differing each other by local Lorentz transformations,
\be
e^{a'} = \Lambda^{a'}{}_b \, e^b,
\ee
which can be written in the very same coordinate system $\{x^\mu\}$. Since a Lorentz transformation does not change the Minkowski metric $\eta_{ab}$, the spacetime metric
\be
g_{\mu \nu} = \eta_{ab} \, e^a{}_\mu e^b{}_\nu,
\ee
is easily seen to be frame independent. As a consequence the Christoffel connection 
\be
\Gammabol^\rho{}_{\mu \nu} = \onehalf g^{\rho \lambda} \left(
\partial_\mu g_{\lambda \nu} + \partial_\nu g_{\lambda \mu} -
\partial_\lambda g_{\mu \nu} \right),
\ee
as well as the corresponding Riemann curvature tensor
\be
\Rbol^\rho{}_{\lambda \mu \nu} = \partial_\mu \Gammabol^\rho{}_{\lambda \nu} -
\partial_\nu \Gammabol^\rho{}_{\lambda \mu} +
\Gammabol^\rho{}_{\eta \mu} \Gammabol^\eta{}_{\lambda \nu} -
\Gammabol^\rho{}_{\eta \nu} \Gammabol^\eta{}_{\lambda \mu},
\ee
are both frame independent. Although frame independent, the Riemann tensor components transform covariantly under the general spacetime coordinate transformation (\ref{difeo}).

%%%%%%%%%%%%%%%%%%%%%%%%%%%%%%%%%%%%%%%%
\section{Linear Approximation of Einstein Equation}
%%%%%%%%%%%%%%%%%%%%%%%%%%%%%%%%%%%%%%%%

%%%%%%%%%%%%%%%%%%%%%%%%%%%%%%%
\subsection{Wave Equation}
%%%%%%%%%%%%%%%%%%%%%%%%%%%%%%%

To study gravitational waves, one has first to obtain the weak field approximation of Einstein equation
\be
\Rbol_{\mu \nu} - \onehalf  \, g_{\mu \nu} \, \Rbol =
\frac{8 \pi G}{c^4} \, \Theta_{\mu \nu},
\ee
with $\Theta_{\mu \nu}$ the source energy-momentum tensor. This is achieved by expanding the metric tensor according to
\be
g_{\mu \nu} = \eta_{\mu \nu} + \varepsilon \, h_{\mt \mu \nu} +
\varepsilon^2 \, h_{\my \mu \nu} + \dots ,
\label{mexpansion}
\ee
where $\varepsilon$ is a small parameter introduced to label the successive orders of this perturbation scheme. At this point, it is important to call the attention to the following fact. When the metric tensor is expanded according to (\ref{mexpansion}), we are automatically assuming that there is a background Minkowskian structure in spacetime, with metric $\eta_{\mu \nu}$. Accordingly, the gravitational waves are interpreted as perturbations
\be
h_{\mu \nu} = \varepsilon \, h_{\mt \mu \nu} +
\varepsilon^2 \, h_{\my \mu \nu} + \dots 
\ee
propagating on that fixed Minkowskian background. This interpretation is consistent with general relativity, as well as with the point of view of field theory, according to which a field always propagates on a background spacetime~\cite{living}.

Assuming expansion (\ref{mexpansion}), the first order Ricci tensor is
\be
\Rbol_{\mt \mu \nu} = \partial_\rho \Gammabol_\mt^\rho{}_{\mu \nu} -
\partial_\nu \Gammabol_\mt^\rho{}_{\mu \rho} \, .
\ee
Using the first order Christoffel connection
\be
\Gammabol_\mt^\rho{}_{\mu \nu} = \onehalf \eta^{\rho \lambda} \left(
\partial_\mu h_{\mt \lambda \nu} + \partial_\nu h_{\mt \lambda \mu} -
\partial_\lambda h_{\mt \mu \nu} \right) , 
\label{Christo}
\ee
the Ricci and scalar curvature tensors become, respectively,
\be
\Rbol_{\mt \mu \nu} = \onehalf \left(\Box h_{\mt \mu \nu} +
\partial_\lambda \partial_\mu h_\mt^\lambda{}_\nu +
\partial_\lambda \partial_\nu h_\mt^\lambda{}_\mu -
\partial_\mu \partial_\nu h_\mt \right)
\ee
and
\be
\Rbol_\mt = \Box h_\mt + \partial_\lambda \partial_\nu h_\mt^{\lambda \nu},
\ee
where $\Box = -\, \eta^{\rho \lambda} \, \partial_\rho \partial_\lambda$ is the flat spacetime d'Alembertian, and $h_\mt = h_\mt^{\lambda}{}_{\lambda}$. In consequence, the first order sourceless gravitational field equation
becomes
\begin{equation}
\Box \left(h_{\mt\mu\nu} - \eta_{\mu\nu}\,h_\mt \right) -
\partial_\mu \partial_\nu h_{\mbox{\tiny{(1)}}} - \eta_{\mu\nu} \, \partial_\alpha
\partial_\beta h_{\mbox{\tiny{(1)}}}^{\alpha\beta} + \partial_\mu \partial_\lambda
h_{\mbox{\tiny{(1)}}}^\lambda{}_\nu + \partial_\nu \partial_\lambda h_{\mbox{\tiny{(1)}}}^\lambda{}_\mu = 0.
\label{eqs4}
\end{equation}

Now, as is well known, wave equation (\ref{eqs4}) is invariant under general spacetime coordinate transformations. Analogously to the electromagnetic wave equation, which is invariant under gauge transformations, the ambiguity of the gravitational wave equation can be removed by choosing a particular class of coordinate system --- or gauge, as it is usually called. The most convenient choice is the class of harmonic coordinate systems, which at first order is fixed by
\be
\eta^{\mu \nu} \, \Gammabol_\mt^\rho{}_{\mu \nu} = 0.
\label{hcc}
\ee
In terms of the metric perturbation, this condition gives
\be
\partial_\rho {h}_\mt^\rho{}_\mu = \onehalf \, \partial_\mu h_\mt.
\label{harmo0}
\ee
In this case, the field equation (\ref{eqs4}) reduces to the relativistic wave equation
\be
\Box \, \phi_{\mbox{\tiny{(1)}}}^{\mu}{}_{\nu} = 0,
\label{we1}
\ee
where we have introduced the new variable
\begin{equation}
\phi_{\mbox{\tiny{(1)}}}^{\mu}{}_{\nu} = h_{\mbox{\tiny{(1)}}}^\mu{}_\nu - \onehalf \, \delta^\mu{}_\nu \, h_{\mbox{\tiny{(1)}}}.
\end{equation}
The harmonic coordinate condition (\ref{harmo0}) can then be expressed in the Lorentz-like form
\be
\partial_\mu  \phi_{\mbox{\tiny{(1)}}}^{\mu}{}_{\nu} = 0.
\label{harmo1}
\ee

%%%%%%%%%%%%%%%%%%%%%%%%%
\subsection{Linear Waves}
\label{TTcs}
%%%%%%%%%%%%%%%%%%%%%%%%%

A plane-wave solution to the relativistic wave equation (\ref{we1}) has the form
\be
\phi_{{\mbox{\tiny{(1)}}}\mu \nu} = A_{{\mbox{\tiny{(1)}}}\mu \nu}  \exp[i k_{\rho} x^\rho],
\label{pw}
\ee
where $A_{{\mbox{\tiny{(1)}}}\mu \nu}=A_{{\mbox{\tiny{(1)}}}\nu \mu}$ is the polarization tensor, and the wave vector $k^\rho$ satisfies
\be
k_{\rho} \, k_{}^\rho = 0.
\ee
The harmonic coordinate condition (\ref{harmo1}), on the other hand, implies
\be
k_\mu \, A_\mt^\mu{}_\nu = 0 \quad \mbox{and} \quad
\partial_\mu h_\mt^\mu{}_\nu = \onehalf \, \partial_\nu h_\mt.
\ee

Still in analogy with the Lorentz gauge in electromagnetism, it is possible to further specialize the harmonic class of coordinates to a particular coordinate system. Once this is done, the coordinate system becomes completely specified, and the components $A_{\mt \mu \nu}$ turn out to represent only physical degrees of freedom. A quite convenient choice is the so called {\it transverse-traceless} coordinate system (or gauge), in which~\cite{schutz}
\be
A_{\mbox{\tiny{(1)}}}^\rho{}_\rho = 0 \quad \mbox{and} \quad
A_{\mbox{\tiny{(1)}}}^\mu{}_\nu \, U_{\mbox{\tiny{(0)}}}^\nu = 0,
\label{TTA}
\ee
with $U_{\mbox{\tiny{(0)}}}^\nu$ an arbitrary, constant four-velocity. Since the trace condition $A_{\mbox{\tiny{(1)}}}^\rho{}_\rho = 0$ implies that $\phi_{{\mbox{\tiny{(1)}}}\mu \nu}=h_{{\mbox{\tiny{(1)}}}\mu \nu}$, we also have
\be
h_{\mbox{\tiny{(1)}}}^\rho{}_\rho = 0, \quad
h_{\mbox{\tiny{(1)}}}^\mu{}_\nu \, U_{\mbox{\tiny{(0)}}}^\nu = 0 \quad \mbox{and} \quad
k_{\mu} \, h_{\mbox{\tiny{(1)}}}^\mu{}_\nu = 0.
\label{TTB}
\ee

Now, although the coordinate system $\{x^\mu\}$ has already been completely specified (the transverse-traceless coordinate system), we still have the freedom to choose different local Lorentz frames $e^a$. Since the metric $g_{\mu \nu} = \eta_{ab} \, e^a{}_\mu e^b{}_\nu$ is invariant under changes of frames, the metric perturbation will also be invariant. In particular, it is always possible to choose a frame fixed at one of the particles, called {\it proper frame}, in which $U_{\mbox{\tiny{(0)}}}^\nu = \delta_0{}^\nu$. In this frame, as can be seen from the second of the Eqs.~(\ref{TTB}),
\be
h_{\mbox{\tiny{(1)}}}^\mu{}_0 = 0
\ee
for all $\mu$. Linear waves satisfying these conditions are usually assumed to represent a plane gravitational wave in the transverse-traceless gauge, propagating in the vacuum with the speed of light. Its physical significance, however, can only be determined by analyzing the energy and momentum it transports.

%%%%%%%%%%%%%%%%%%%%%
\subsection{Energy and Momentum Transported by Linear Waves}
%%%%%%%%%%%%%%%%%%%%%

The energy-momentum tensor of any matter (or source) field $\psi$ is always proportional to the functional derivative of the corresponding Lagrangian with respect to the spacetime metric. Since a derivative with respect to the metric does not change the order of the Lagrangian in the matter field $\psi$, both the Lagrangian and the energy-momentum tensor will be of the same order in the field variables $\psi$. For example, since Maxwell's Lagrangian is quadratic in the electromagnetic field, the corresponding energy-momentum tensor will also be quadratic. Now, as is well known, the gravitational field is itself source of gravitation. This means that the gravitational energy-momentum density must appear explicitly in the gravitational field equation. In fact, the correct form of the wave equation (\ref{we1}) is
\be
\Box \, \phi_{\mt \mu \nu} - \frac{16 \pi G}{c^4} \, t_{\mt \mu \nu}  = 0,
\label{we1bis}
\ee
with $t_{\mt \mu \nu}$ the first-order gravitational energy-momentum pseudotensor. As a consequence, {\em in the linear ap\-prox\-imation}, the gravitational energy-momentum density is also restricted to be linear. Since the energy-momentum density is at least quadratic in the field variable, it vanishes in the linear approximation, yielding the wave equation~(\ref{we1}). Of course, if this energy-momentum density is used to calculate the energy and momentum transported by linear gravitational waves, the result will obviously be that they do not carry neither energy nor momentum.

The consistency of this result can be verified by analyzing the generation of linear waves. In the presence of a source, the first-order field equation reads
\be
\Box \, \phi_{\mt \mu \nu} = \frac{16 \pi G}{c^4} \; \Theta_{\mt \mu \nu},
\label{fo}
\ee
with $\Theta_{\mt \mu \nu}$ the first-order source energy-momentum tensor. One solution is the retarded potential
\be
\phi_{{\mbox{\tiny{(1)}}}\mu \nu}(\vec{x}, t) = \frac{4 G}{c^4} \int 
\frac{d^3x'}{| \vec{x} - \vec{x}' |} \; \Theta_{{\mbox{\tiny{(1)}}}\mu \nu}(\vec{x}', t'),
\ee
where the gravitational source is considered in the retarded time $t' = t - | \vec{x} - \vec{x}' |/c$. Now, as a consequence of the coordinate condition (\ref{harmo1}), it is easy to see that
\be
\partial_\nu {\Theta}_{{\mbox{\tiny{(1)}}}\mu}{}^\nu = 0.
\label{focl}
\ee
Instead of the usual covariant derivative, $\Theta_{{\mbox{\tiny{(1)}}}\mu}{}^\nu$ is conserved with an ordinary derivative at the first order. Since this is a true conservation law, in the sense that it leads to a time conserved {\it charge} --- which in this case represents the source energy and momentum --- we can conclude that in the linear approximation a mechanical system cannot lose energy and momentum in the form of gravitational radiation. This is consistent with the fact that linear gravitational waves do not transport energy and momentum.

It is important to observe that, in the case of the electromagnetic field, the linearity of the wave equation does not restrict the electromagnetic energy-momentum tensor to be linear. To understand this point, let us write Maxwell equation in the form
\be
\partial_\mu F^{\mu \nu} - j^\nu = J^\nu,
\ee
where we have introduced the electromagnetic self-current $j^\nu$. As is well known, the linearity of electromagnetism restricts the electromagnetic self-current to be linear, and consequently to vanish: $j^\nu = 0$. In fact, the electromagnetic wave is well known not to transport its own charge --- that is, electric charge. The consistency of this result can be verified by observing that, in the electromagnetic case, the conservation law corresponding to (\ref{focl}) is the conservation of the electric four current, $\partial_\mu J^\mu = 0$. Such conservation law precludes the electromagnetic wave to transport electric charge, but not energy and momentum. This is a fundamental difference between electromagnetic and {\it linear} gravitational waves.

%%%%%%%%%%%%%%%%%%%%%%%%%
\section{A Glimpse on Nonlinear Waves}
%%%%%%%%%%%%%%%%%%%%%%%%%

At the second order of an iterated perturbation scheme~\cite{living}, the harmonic coordinate condition reads
\be
\eta^{\mu \nu} \, \Gamma_\my^\rho{}_{\mu \nu} = 0.
\ee
This is equivalent to
\be
\partial_\mu \phi_{\mbox{\tiny{(2)}}}^{\mu}{}_{\nu} = 0,
\label{harmo2}
\ee
where
\begin{equation}
\phi_{\mbox{\tiny{(2)}}}^{\mu}{}_{\nu} = h_{\mbox{\tiny{(2)}}}^\mu{}_\nu - \onehalf \delta^\mu{}_\nu \, h_{\mbox{\tiny{(2)}}},
\end{equation}
with $h_\my = h_\my^\rho{}_\rho$. In these coordinates, the second order gravitational field equation can be written in the form
\be
\Box \, \phi_{\my}^{\mu}{}_{\nu} - \frac{16 \pi G}{c^4} \, t_{\my}^{\mu}{}_{\nu} =
\frac{16 \pi G}{c^4} \; \Theta_{\my}^{\mu}{}_{\nu},
\label{sofe1}
\ee
where $t_{\my}^{\mu}{}_{\nu} \equiv t_{\my}^{\mu}{}_{\nu}(h_\mt, h_\mt)$ represents all terms coming from the left-hand side of Einstein equation, in addition to the d'Alembertian term. It can be interpreted as the energy-momentum pseudotensor of the gravitational field.

Now, as can be seen from Eqs.~(\ref{harmo2}) and (\ref{sofe1}), the second-order total energy-momentum tensor is conserved:
\be
\partial_\mu \left[t_{\my}^{\mu}{}_{\nu} + \Theta_{\my}^{\mu}{}_{\nu} \right] = 0.
\ee
The source energy-momentum tensor, on the other hand, as determined by the second order Bianchi identity, is conserved in the covariant sense:
\be
\nabla_\mu \Theta_{\my}^{\mu}{}_{\nu} \equiv
\partial_\mu \Theta_{\my}^{\mu}{}_{\nu} +
\Gammabol_\mt^\mu{}_{\rho \mu} \, \Theta_{\mt}^{\rho}{}_{\nu} -
\Gammabol_\mt^\rho{}_{\nu \mu} \, \Theta_{\mt}^{\mu}{}_{\rho} = 0.
\ee
At the second order, therefore, the source energy-momentum tensor is not conserved. In fact, the above covariant conservation law is not a true conservation law, but simply an identity governing the exchange of energy and momentum between gravitation and matter~\cite{kopo}. As a consequence, in contrast to what happens at the first order, at the second order a mechanical system can lose energy in the form of gravitational waves.

Far away from the sources, the second order gravitational waves are governed by the sourceless version of the wave equation (\ref{sofe1}), which can be written in the form
\be
\Box \, \phi_{\my}^{\mu}{}_{\nu} = N^{\mu}{}_{\nu}(h_\mt, h_\mt),
\label{sofeless}
\ee
with $N^{\mu}{}_{\nu}(h_\mt, h_\mt) = (16 \pi G/c^4) \, t_{\my}^{\mu}{}_{\nu}$. Since $h_{{\mbox{\tiny{(1)}}}\mu \nu}$ is known, the problem reduces to that of solving a wave equation whose source is given.

%%%%%%%%%%%%%%%%%%%%%%%%%%%%%%
\subsection{Emission of Gravitational Waves}
%%%%%%%%%%%%%%%%%%%%%%%%%%%%%%

Considering the wave zone, that is, distances much larger than the dimensions of the source, the power emitted per unit solid angle in a given direction\footnote{Middle letters ($i, j, k, \dots = 1, 2, 3$) of the  Latin alphabet will be used to denote space indices.} $\hat{x} = \vec{x}/r$, with $r = |\vec{x}|$, is given by~\cite{weinberg}
\be
\frac{dP}{d\Omega} = r^2 \, \hat{x}^i  < t^{i0} >,
\label{power}
\ee
where $< t^{i0} >$ is the energy flux averaged over a spacetime dimension large compared with $1/\omega$, with $\omega$ the frequency of the wave. Now, in a perturbation scheme, different orders of the gravitational field expansion will give rise to different-order contributions to the power emitted. This means that we have to expand
\be
P = \varepsilon \, P_{\mbox{\tiny{(1)}}} + \varepsilon^2 \, P_{\mbox{\tiny{(2)}}} + \dots \, .
\ee
At the first order, $t_{{\mbox{\tiny{(1)}}}}^{\mu \nu}$ vanishes, and the corresponding power emitted will also vanish:
\be
\frac{dP_{{\mbox{\tiny{(1)}}}}}{d\Omega} \equiv r^2 \, \hat{x}^i  < t_{{\mbox{\tiny{(1)}}}}^{i0} > = 0.
\label{power1}
\ee
This is consistent with the conservation law (\ref{focl}), and also with the fact that first-order (that is, linear) gravitational waves do not transport energy and momentum.

At the second order, the power emitted by a mechanical system will be
\be
\frac{dP_{{\mbox{\tiny{(2)}}}}}{d\Omega} = r^2 \, \hat{x}^i  < t_{{\mbox{\tiny{(2)}}}}^{i0} >.
\label{power2}
\ee
As already discussed, the energy-momentum pseudotensor $t_{{\mbox{\tiny{(2)}}}}^{i0}$ depends only on the first-order solution. This means that Eq.~(\ref{power2}) coincides with the usual expression found in all textbooks. It yields, therefore, the well known expression for the power emitted by a mechanical source, and in particular the usual quadrupole radiation formula. There is a difference, though: as a second-order effect, the energy and momentum emitted cannot be transported away by a linear wave, as is usually assumed, but only by a nonlinear gravitational wave.

%%%%%%%%%%%%%%%%%%%%%%%%%%%%%%
\section{Effects on Free Particles}
%%%%%%%%%%%%%%%%%%%%%%%%%%%%%%

%%%%%%%%%%%%%%%%%%%%%%%%%%%%%%
\subsection{The Geodesic Deviation Equation}
%%%%%%%%%%%%%%%%%%%%%%%%%%%%%%

Let us consider, as usual, two nearby particles separated by the four-vector $\xi^\alpha$. In the context of general relativity, this vector obeys the geodesic deviation equation
\be
\nabol_U \, \nabol_U \xi^\alpha = \Rbol^\alpha{}_{\mu \nu \beta} \,
U^\mu \, U^\nu \, \xi^\beta,
\label{gde0}
\ee
where $U^\mu = dx^\mu/ds$, with $ds = g_{\mu \nu} \, dx^\mu dx^\nu$, is the four-velocity of the particles. Now, each order of the gravitational field expansion 
\be
\Rbol^\alpha{}_{\mu \nu \beta} = \varepsilon \, \Rbol_{\mbox{\tiny{(1)}}}^\alpha{}_{\mu \nu \beta} +
\varepsilon^2 \, \Rbol_{\mbox{\tiny{(2)}}}^\alpha{}_{\mu \nu \beta} + \dots \, ,
\ee
which follows naturally from (\ref{mexpansion}), will give rise to a different contribution to $\xi^\alpha$. For consistence reasons, therefore, this vector must also be expanded:
\be
\xi^\alpha = \xi_{\mbox{\tiny{(0)}}}^\alpha + \varepsilon \, \xi_{\mbox{\tiny{(1)}}}^\alpha +
\varepsilon^2 \, \xi_{\mbox{\tiny{(2)}}}^\alpha + \dots \; .
\ee
In this expansion, $\xi_{\mbox{\tiny{(0)}}}^\alpha$ represents the initial, that is, undisturbed separation between the particles. Of course, as the four-velocity $U^\mu$ depends on the gravitational field, it must also be expanded. We then write
\be
U^\mu = U_{\mbox{\tiny{(0)}}}^\mu + \varepsilon \, U_{\mbox{\tiny{(1)}}}^\mu +
\varepsilon^2 \, U_{\mbox{\tiny{(2)}}}^\mu + \dots \; ,
\ee
where $U_{\mbox{\tiny{(0)}}}^\mu$ is a constant arbitrary four-velocity, which depends on the choice of the initial condition --- or equivalently, on the choice of the local Lorentz frame from which the phenomenon will be observed and measured. Using then the freedom to choose this frame (see section \ref{TTcs}), we can choose a frame fixed at one of the particles --- called {\it proper frame}. At the lowest order, therefore, the four-velocity $U_{\mbox{\tiny{(0)}}}^\mu$ can be expressed in terms of the observer proper time $s_{\mbox{\tiny{(0)}}}$, that is, $U_{\mbox{\tiny{(0)}}}^\mu = dx^\mu/ds_{\mbox{\tiny{(0)}}}$, where
\be
ds_{\mbox{\tiny{(0)}}}^2 = \eta_{\mu \nu} \, dx^\mu dx^\nu
\label{ds0}
\ee
is the flat spacetime quadratic interval. Since, in the proper reference frame, the particles are initially at rest, we have that
\be
U_{\mbox{\tiny{(0)}}}^\mu \equiv \delta^\mu{}_0 = (1, 0, 0, 0),
\label{U0}
\ee
which means that, in this frame, the proper time $s_{\mbox{\tiny{(0)}}}$ coincides with the coordinate $x^0$~\cite{mtw}.

Before proceeding further, it is crucial to observe that the geodesic deviation equation (\ref{gde0}) is covariant under general coordinate transformations. In  consequence, the equations emerging at each order of the perturbation scheme will also be covariant up to the corresponding order. Due to this covariance, one can either solve the equation in a general coordinate system and then specialize to the transverse-traceless coordinates, which are the coordinates used to describe the gravitational wave, or specialize first to these coordinates and then solve the ensuing differential equations. Here, we will adopt the first alternative, which we consider far more convenient.

%%%%%%%%%%%%%%%%%%%%%%%%%%%%%%
\subsection{Linear Effects}
%%%%%%%%%%%%%%%%%%%%%%%%%%%%%%

At the lowest order, the geodesic deviation equation (\ref{gde0}) becomes
\be
\frac{d^2 \xi_{\mbox{\tiny{(0)}}}^\alpha}{ds_{\mbox{\tiny{(0)}}}^2} = 0.
\label{gde1}
\ee
This is an expected result, in the sense that $\xi_{\mbox{\tiny{(0)}}}^\alpha$ is simply the undisturbed separation between the particles. Without loss of generality, we can take the solution to be $\xi_{\mbox{\tiny{(0)}}}^\alpha$ = constant. At the next order, the geodesic deviation equation becomes
\be
\frac{d^2 \xi_{\mbox{\tiny{(1)}}}^\alpha}{ds_{\mbox{\tiny{(0)}}}^2} +
U^\rho_{\mbox{\tiny{(0)}}} \, \partial_\rho \left(\Gammabol_{{\mbox{\tiny{(1)}}}}^\alpha{}_{\beta \gamma} \, U^\gamma_{\mbox{\tiny{(0)}}} \right) \xi_{\mbox{\tiny{(0)}}}^\beta  =
\Rbol_{\mbox{\tiny{(1)}}}^\alpha{}_{\mu \nu \beta} \, U^\mu_{\mbox{\tiny{(0)}}} \, U^\nu_{\mbox{\tiny{(0)}}} \; \xi_{\mbox{\tiny{(0)}}}^\beta.
\label{gde01}
\ee
Substituting $U^\mu_{\mbox{\tiny{(0)}}}$ as given by Eq.~(\ref{U0}), we get
\be
\frac{d^2 \xi_{\mbox{\tiny{(1)}}}^\alpha}{ds_{\mbox{\tiny{(0)}}}^2} + \partial_0 \Gammabol_{{\mbox{\tiny{(1)}}}}^\alpha{}_{\beta 0} \; \xi_{\mbox{\tiny{(0)}}}^\beta  = \Rbol_{\mbox{\tiny{(1)}}}^\alpha{}_{00\beta} \; \xi_{\mbox{\tiny{(0)}}}^\beta.
\label{gde2}
\ee
Using then the first-order Riemann tensor
\be
\Rbol_{\mbox{\tiny{(1)}}}^\alpha{}_{\mu \nu \beta} =
\partial_\nu \Gammabol_{\mt}^\alpha{}_{\mu \beta} -
\partial_\beta \Gammabol_{{\mbox{\tiny{(1)}}}}^\alpha{}_{\mu \nu},
\label{R1}
\ee
it reduces to
\be
\frac{d^2 \xi_{\mbox{\tiny{(1)}}}^\alpha}{ds_{\mbox{\tiny{(0)}}}^2} + \partial_0 \Gammabol_{{\mbox{\tiny{(1)}}}}^\alpha{}_{\beta 0} \; \xi_{\mbox{\tiny{(0)}}}^\beta  = \left( \partial_0 \Gammabol_{{\mbox{\tiny{(1)}}}}^\alpha{}_{\beta 0} -
\partial_\beta \Gammabol_{{\mbox{\tiny{(1)}}}}^\alpha{}_{0 0} \right) \xi_{\mbox{\tiny{(0)}}}^\beta.
\label{gde2bis}
\ee
Canceling $\partial_0 \Gammabol_{{\mbox{\tiny{(1)}}}}^\alpha{}_{\beta 0} \; \xi_{\mbox{\tiny{(0)}}}^\beta$ on both sides, we get
\be
\frac{d^2 \xi_{\mbox{\tiny{(1)}}}^\alpha}{ds_{\mbox{\tiny{(0)}}}^2} = - \, \partial_\beta \Gammabol_{{\mbox{\tiny{(1)}}}}^\alpha{}_{0 0} \; \xi_{\mbox{\tiny{(0)}}}^\beta,
\label{gde3}
\ee
where
\be
\Gammabol_{\mt}^\alpha{}_{0 0} = \onehalf \, \eta^{\alpha \rho} \left( 2 \; \partial_0 h_{\mt\rho 0} -
\partial_\rho h_{\mt 00} \right).
\ee
Specializing now to the transverse-traceless coordinate system, where the components $h_{\mt \rho 0}$ vanish identically, we obtain
\be
\frac{d^2 \xi_{\mt}^\alpha}{ds_{\mbox{\tiny{(0)}}}^2} = 0.
\label{gde4}
\ee
This means that, in the linear approximation, the particles are not affected by gravitational waves.

For the sake of completeness, let us consider also the other usual approach to the same question, which consists in calculating the {\it proper distance} between two nearby particles. Such a distance is defined by
\be
\Delta l = \int |ds^2|^{1/2},
\ee
with $ds^2 = g_{\mu \nu} \, dx^\mu dx^\nu$. Due to the fact that the gravitational wave solution is written in a very specific coordinate system, this computation has to be done either in this specific coordinate system or using a coordinate--independent formulation. Here, we will follow the second alternative.\footnote{In all approaches found in the literature, this computation is made in a coordinate--dependent formulation. However, these approaches are not consistent, in the sense that different coordinate systems are used to describe the wave and the position of the particles. In fact, whereas the gravitational wave is described in transverse-traceless coordinates, the position of the particles are usually specified in Cartesian coordinates.} 
Using then the metric expansion (\ref{mexpansion}), and keeping terms up to first order, we obtain
\be
ds^2 = ds_{\mbox{\tiny{(0)}}}^2 +
\varepsilon \, h_{{\mbox{\tiny{(1)}}} \mu \nu} \, dx^\mu dx^\nu,
\ee
with $ds_{\mbox{\tiny{(0)}}}^2$ given by Eq.~(\ref{ds0}). Choosing again the proper reference frame fixed at one of the particles, we can write
\be
ds^2 = \left( 1 + \varepsilon \, h_{{\mbox{\tiny{(1)}}} \mu \nu} \, U_{\mbox{\tiny{(0)}}}^\mu \, U_{\mbox{\tiny{(0)}}}^\nu \right) ds_{\mbox{\tiny{(0)}}}^2,
\ee
where $U_{\mbox{\tiny{(0)}}}^\mu = dx^\mu/ds_{\mbox{\tiny{(0)}}}$. However, in the transverse-traceless gauge,  
\[
h_{{\mbox{\tiny{(1)}}} \mu \nu} \, U_{\mbox{\tiny{(0)}}}^\mu = 0.
\]
Up to first order, therefore, the proper distance between two nearby particles is simply
\be
\Delta l = \int |ds_{\mbox{\tiny{(0)}}}^2|^{1/2},
\ee
which means that it is not changed by a first-order gravitational wave. Only at the next order, where nonlinear effects will be in action, will the proper distance be affected by a gravitational wave. 

Although some of the components of the first-order curvature tensor $\Rbol_{\mbox{\tiny{(1)}}}^\alpha{}_{\mu \nu \beta}$ do not vanish in the transverse-traceless coordinate system, as we have seen, they are unable to induce any movement on free particles. Since these components are different in different coordinate systems, it is difficult to get some insight on their meaning. A better way to understand the effects of curvature is to inspect the invariants~\cite{cw} constructed out of $\Rbol_{\mbox{\tiny{(1)}}}^\alpha{}_{\mu \nu \beta}$, as for example the Kretschmann
\[
{\mathcal I}{\mbox{\tiny{(2)}}} = \Rbol_{\mbox{\tiny{(1)}}}^{\alpha \beta \mu \nu} \, \Rbol_{{\mbox{\tiny{(1)}}}\alpha \beta \mu \nu}
\]
and the pseudo-scalar
\[
{\mathcal I}'{\mbox{\tiny{(2)}}} = \star\Rbol_{\mbox{\tiny{(1)}}}^{\alpha \beta \mu \nu} \, \Rbol_{{\mbox{\tiny{(1)}}}\alpha \beta \mu \nu}  \equiv
\onehalf \epsilon^{\alpha \beta \lambda \rho} \, \Rbol_{{\mbox{\tiny{(1)}}}\lambda \rho}{}^{\mu \nu} \, \Rbol_{{\mbox{\tiny{(1)}}}\alpha \beta \mu \nu},
\]
with $\star$ denoting the Hodge dual. As an easy calculation shows, both invariants vanish everywhere for the linear solution (\ref{pw}). It is not surprising, therefore, that such waves are unable to impart energy and momentum to free particles.

%%%%%%%%%%%%%%%%%%%%%
\subsection{Some Comments on the Usual Procedure}
%%%%%%%%%%%%%%%%%%%%%

In all approaches found in the literature, the effect produced by linear gravitational waves on free particles is computed by first specializing the geodesic deviation equation to the transverse-traceless coordinates, and then by solving the corresponding differential equations. Of course, since the coordinate system is fixed from the very beginning, these approaches are not manifestly covariant. Moreover, all of them present some unclear points, which we pass to discuss.

It is usually claimed that, by choosing a frame fixed at one of the particles, the Christoffel connection can be made to vanish, not only  at one point, but along the whole world-line of a comoving observer fixed at this frame.\footnote{We notice in passing that there is an apparent problem with this claim: as we have discussed in section \ref{basic}, the Christoffel coefficients do not depend on the frame, and hence cannot be made to vanish by a choice of a frame. In order to make a connection to vanish along a curve, it is necessary to transform to a locally inertial coordinate system, sometimes called Galilean coordinates~\cite{landau}. However, this is not possible here because the coordinate system has already been completely specified. Remember that the gravitational waves are described in the transverse-traceless coordinate system. In addition, as a simple inspection shows, not all components of the Christoffel connection (\ref{Christo}) vanish in the transverse-traceless coordinates, which means that this coordinate system is not Galilean.} In other words, it is usually claimed that~\cite{mtw}
\be
\frac{d \Gammabol_{\mbox{\tiny{(1)}}}^\alpha{}_{\beta \gamma}}{ds_{\mbox{\tiny{(0)}}}} \equiv \partial_0 \Gammabol_{\mbox{\tiny{(1)}}}^\alpha{}_{\beta \gamma} = 0,
\label{dg0}
\ee
where use has been made of the fact that, in this frame, the proper time $s_{\mbox{\tiny{(0)}}}$ coincides with the coordinate $x^0$. This result is then used to eliminate the connection-term from the left-hand side of Eq.~(\ref{gde2bis}), {\it but not the same connection-term appearing in the right-hand side of the equation}~\cite{leclerc}. The argument used to justify this procedure is that, since $\Rbol_{\mbox{\tiny{(1)}}}^\alpha{}_{\mu \nu \beta}$ is a tensor, it can be calculated in any coordinate system, not necessarily in the transverse-traceless coordinates. This argument is, however, misleading. The reason is that, once the coordinate system has been completely specified, the equations become not manifestly covariant, and consequently any argument based on covariance cannot be applied. Actually, consistency reasons do require that, since the geodesic deviation equation is written in the transverse-traceless coordinates, the Riemann tensor components on the right-hand side of this equation {\em must necessarily be expressed in the same coordinate system}. As a consequence, if the connection-term in the left-hand side of Eq.~(\ref{gde2bis}) is found to vanish, the connection-term appearing in the right-hand side of the same equation must also be assumed to vanish.\footnote{Observe that this would not mean that all components of the Riemann tensor vanish: only its projection along the particle four-velocity would vanish, that is,
$
\Rbol_{\mbox{\tiny{(1)}}}^\alpha{}_{\mu \nu \beta} \, U^\mu_{\mbox{\tiny{(0)}}} \, U^\nu_{\mbox{\tiny{(0)}}} \equiv
\Rbol_{\mbox{\tiny{(1)}}}^\alpha{}_{0 0 \beta} = 0.
$
As can be easily verified, other components of the Riemann tensor would still be non-vanishing to comply with its covariant behavior.} Furthermore, observe that, since the left-hand side as a whole is also a tensor, the use of the same criterion should preclude the elimination of its connection term. 

The important point, we repeat, is that arguments based on covariance cannot be used when the coordinate system has been completely determined. Notice, for example, that in the transverse-traceless coordinate system, we obtain from Eq.~(\ref{R1}) the relation
\be
\Rbol_{\mbox{\tiny{(1)}}}^\alpha{}_{\mu \nu \beta} \, U^\mu_{\mbox{\tiny{(0)}}} \, U^\nu_{\mbox{\tiny{(0)}}} =
U^\rho_{\mbox{\tiny{(0)}}} \, \partial_\rho \left(\Gammabol_{{\mbox{\tiny{(1)}}}}^\alpha{}_{\beta \gamma} \, U^\gamma_{\mbox{\tiny{(0)}}} \right).
\ee
Of course, this identity is not inconsistent with covariance because it is valid only in that specific coordinate system. Therefore, when the geodesic deviation equation (\ref{gde2}) is restricted to the transverse-traceless coordinates, the above two terms can be consistently canceled from the geodesic deviation equation (\ref{gde2}), yielding the equation of motion (\ref{gde4}).

%%%%%%%%%%%%%%%%%%%%%
\section{Final Remarks}
%%%%%%%%%%%%%%%%%%%%%

Considering that the sources of gravitational waves are at enormous distances from Earth, the amplitude of these waves when reaching a detector on Earth are expected to be so small that the linearized theory is usually assumed to be enough to accurately describe them. In other words, to comply with the idea that the radiative solutions to the gravitational field equation should satisfy a linear wave-equation, these waves are usually assumed to carry not enough energy and momentum to affect their own propagation.\footnote{This is a commonplace in gravitational wave theory. For a textbook reference, see \cite{weinberg}.} However, these assumptions are clearly misleading. The basic reason is that this is not a matter of approximation, but a conceptual question. Either a gravitational wave does or does not carry energy: if it carries, it cannot satisfy a linear equation. It is, therefore, {\em conceptually} unacceptable to assume that a gravitational wave satisfying a linear equation is able to transport energy and momentum. Only a nonlinear gravitational wave is able to do it. The above assumption would correspond to assume that, if a solitary-wave solution to the Korteweg-de Vries equation had a small-enough amplitude, its evolution could be accurately described by a linear equation. Needless to say, this is plainly wrong.

The problem of defining an energy-momentum density for the gravitational field has a close analogy with the problem of defining a gauge current for the Yang-Mills field, whose sourceless field equation is~\cite{itzu}
\be
\partial_\mu F^{A \mu \nu} + j^{A \nu} = 0.
\label{ym}
\ee
The piece
\be
j^{A \nu} = - f^A{}_{B C} \, A^B{}_\mu \, F^{C \mu \nu}
\ee
represents the gauge pseudocurrent, which is present only for non-Abelian gauge groups (for which $f^A{}_{B C} \ne 0$).  Due to the explicit presence of the connection $A^B{}_\mu$, this current is clearly not gauge covariant, in analogy with the gravitational energy-momentum pseudotensor, which is not covariant under general coordinate transformations. A fundamental characteristic of the Yang-Mills field is that it carries its own charge, as for example the {\it color} charge carried by gluons in chromodynamics. This simple property precludes the existence of Yang-Mills linear waves. In fact, linear Yang-Mills waves --- that is, solutions of the linearized Yang-Mills equation --- are waves that do not carry their own charge, a property also revealed by the fact that, in the linear approximation, the pseudocurrent $j^{A \nu}$ vanishes. The Yang--Mills field, therefore, is essentially nonlinear --- otherwise it is not a Yang-Mills field.

Let us consider now the gravitational field, whose sourceless field equation, in the so called potential form, is written as~\cite{moller}
\begin{equation}
\partial_\mu (\sqrt{-g} \, S_\lambda{}^{\rho \mu}) -
\frac{8 \pi G}{c^4} \, \sqrt{-g} \, t_\lambda{}^\rho = 0,
\label{eqs1bis}
\end{equation}
where $g = \det(g_{\mu \nu})$ and $S_\lambda{}^{\rho \mu} = - S_\lambda{}^{\mu \rho}$ is the superpotential. This equation presents a structure similar to the Yang-Mills equation (\ref{ym}), with the gravitational energy-momentum pseudo-density $({8 \pi G}/{c^4}) \, h t_\lambda{}^\rho$ playing the role of self-current. This similarity comes from the fact that, analogously to Yang-Mills fields, the gravitational field also carries its own charge --- in the case, energy and momentum. Because the energy-momentum pseudotensor is at least quadratic in the field variables~\cite{living}, it vanishes in the linear approximation. Up to first order, therefore, the sourceless field equation (\ref{eqs1bis}) becomes the  field equation (\ref{eqs4}), which in harmonic coordinates reduces to
\be
\Box \, h_{\mt \mu \nu} = 0.
\label{**}
\ee
In this form, it is similar to Maxwell's equation in the Lorentz gauge. This could be interpreted as an indication that, in this approximation, the gravitational field loses its similarity to Yang-Mills fields, becoming similar to the electromagnetic field, whose self-current also vanishes. However, for the same reason a Yang-Mills field must be nonlinear to carry its own charge, a gravitational wave must be nonlinear to transport energy and momentum. Otherwise,  it is not a gravitational wave. It should be remarked that, besides providing the dispersion relation, the solution $h_{\mt \mu \nu}$ of the linear equation (\ref{**}) is physically meaningful,  in the sense that the higher-order solutions depend on it. Nevertheless,  it does not represent by itself a gravitational wave which, like a Yang-Mills propagating field, is a strictly nonlinear phenomenon.\footnote{It should be remarked that gravitation has a physically meaningful linear (static) limit, known as the Newtonian limit. In that limit, although one usually talks about ``gravitational field'', there is actually no field in the usual sense. In fact, in Newtonian gravity the interaction is instantaneous, and there is no energy density associated with gravitation: the total energy of any physical system includes only the kinetic and the potential energies of its components, but not the energy of gravitation itself. Furthermore, it is not simply a weak-field limit, but a completely different theory. Observe, for example, that the local Poincar\'e symmetry of general relativity is contracted to Galilei, the symmetry group of Newtonian gravity. In other words, Newtonian gravity is not just a weak-field approximation, but a contraction limit~\cite{IW} of general relativity.}

Summing up, although the use of a perturbative analysis could suggest the existence of linear gravitational waves, physical reasons show that a gravitational wave is essentially a nonlinear phenomenon. Of course, any mechanical system --- like a binary pulsar, for example --- can lose energy in the form of gravitational radiation. The usual expression for the power emitted by a mechanical source, and in particular the quadrupole radiation formula, give a correct account of this energy. What our results say is simply that this is a second-order effect, and as such the energy and momentum cannot be transported away by a linear wave, but by a solution of a nonlinear wave-equation. At the lowest order, it is given by the second order wave equation
\[
\Box \, \phi_{\my}^{\mu}{}_{\nu} = N^{\mu}{}_{\nu}(h_\mt, h_\mt).
\]
The traveling wave solution to this equation, therefore, represent the physical gravitational wave in the sense that it is able to transport energy and momentum, and consequently to be detected. It is the wave to be searched when looking for gravitational waves. An analysis of these waves, therefore, as well as of their effects on a test particles, turns out to be crucial for determining their signature on gravitational wave detectors.

%%%%%%%%%%%%%%%%%%%%%%%%%%
\section*{Acknowledgments}
The authors would like to thank Yu.\ Obukhov and G.\ Rubilar for useful discussions. They would like to thank also FAPESP, CNPq and CAPES for partial financial support.

%%%%%%%%%%%%%%%%%%%%%%%%%%%

\end{document}